\documentclass[12pt]{iopart}

\usepackage{graphicx}

\begin{document}

\title{Observation of phonons with resonant inelastic x-ray scattering}

\author{H Yava\c{s}$^{1,2}$\footnote{Present address: Deutsches Elektronen-Synchrotron DESY, D-22603 Hamburg, Germany}, M van Veenendaal$^{1,3}$, J van den Brink$^4$, L J P Ament$^4$, A Alatas$^1$, B M Leu$^1$, M-O Apostu$^5$, N Wizent$^{6,7}$, G Behr$^6$, W Sturhahn$^1$\footnote{Present address: Jet Propulsion Laboratory, California Institute of Technology, Pasadena, CA 91109, USA}, H Sinn$^8$ and E E Alp$^1$}

\address{$^1$Advanced Photon Source, Argonne National Laboratory, Argonne, IL 60439, USA}
\address{$^2$Department of Geology, University of Illinois, Urbana, IL 61801, USA}
\address{$^3$Department of Physics, Northern Illinois University, De Kalb, IL 60115, USA}
\address{$^4$Institute-Lorentz for Theoretical Physics, Universiteit Leiden, P.O. Box 9506, 2300 RA Leiden, The Netherlands}
\address{$^5$Faculty of Chemistry, "Al. I. Cuza" University, 11Carol-I, 700506 Iasi-Romania}
\address{$^6$Leibniz-Institut f\"{u}r Festk\"{o}rper- und Werkstoffforschung (IFW) Dresden, D-01171, Dresden Germany}
\address{$^7$Kirchhoff-Institut f\"{u}r Physik, D-69120 Heidelberg Germany}
\address{$^8$DESY, Hasylab, Notkestrasse 85, 22607 Hamburg, Germany}
\ead{hasan.yavas@desy.de}

\begin{abstract}
Phonons, the quantum mechanical representation of lattice vibrations, and their coupling to the electronic degrees of freedom are important for understanding thermal and electric properties of materials. For the first time, phonons have been measured using resonant inelastic x-ray scattering (RIXS) across the Cu $K$-edge in cupric oxide (CuO). Analyzing these spectra using an ultra-short core-hole lifetime approximation and exact diagonalization techniques, we can explain the essential inelastic features. The relative spectral intensities are related to the electron-phonon coupling strengths. 
\end{abstract}

\maketitle

\section{Introduction}

It is generally accepted that the retardation of the Coulomb interaction due to the coupling between electrons and phonons is the mechanism for conventional BCS superconductivity. On the other hand, despite over two decades of unceasing research efforts, the mechanism for high-$T_c$ superconductivity in cuprates remains evasive. Although it has been suggested that electron-phonon coupling could be a key towards understanding this intriguing phenomenon \cite{lanzara01,Cuk04,Devereaux04,reznik06}, the topic is still being heavily debated \cite{Johnson01,giustino08,Dahm09,reznik08}. Whereas information about the phonon states can be obtained through inelastic neutron and x-ray scattering or Raman spectroscopy, the study of the coupling between electrons and phonons is more elusive. A number of experimental techniques provide an indirect measure of the electron-phonon coupling. For example, in angle-resolved photo-emission spectra of high-$T_c$ cuprates, kinks observed in the dispersion of the electron bands have been attributed to strong electron-phonon coupling. However, since the observed spectral features are dominated by electronic excitations, it is difficult to unambiguously assign the origin of these kinks to phonons rather than, for example, spin excitations \cite{Johnson01,giustino08,Dahm09}.  Point-contact spectroscopy \cite{Jansen80}, on the other hand, is sensitive to interaction of electrons with elementary excitations including phonons; however, this technique is not momentum resolved. In addition, the spectra strongly depend on the quality of the tunnel barrier, transmission matrix elements, and inelastic scattering in the barrier region. Resonant Inelastic X-ray Scattering (RIXS) has been shown to be sensitive to phonon excitations, but until now they have only been observed in combination with electronic excitations \cite{Hancock10}.

Here we demonstrate that RIXS can probe pure phonon excitations through the core-electron excitation, that provide element-specific and momentum-resolved information on the coupling between phonons and electrons, but that separates the phonon and the electron degrees of freedom in the final states. To achieve this, we employed RIXS at the transition-metal $K$-edge. The underlying physical mechanism is as follows. The incoming x-ray excites an electron from the deep-lying $1s$ core level into the valence band consisting of $3d$ and $4p$ states through quadrupolar and dipolar transitions, respectively. At the site where the absorption takes place, this excitation creates a sudden change in charge density, to which the lattice responds through the electron-phonon coupling. Since the deexcitation removes the electronic excitation created in the absorption, final states with only phonon excitations can be reached. Additionally, the very short core-hole lifetime prevents a complete relaxation of the lattice within the time scale of the resonant scattering process.

\section{Experiment}

\begin{figure}
\centering
\includegraphics[scale=0.67]{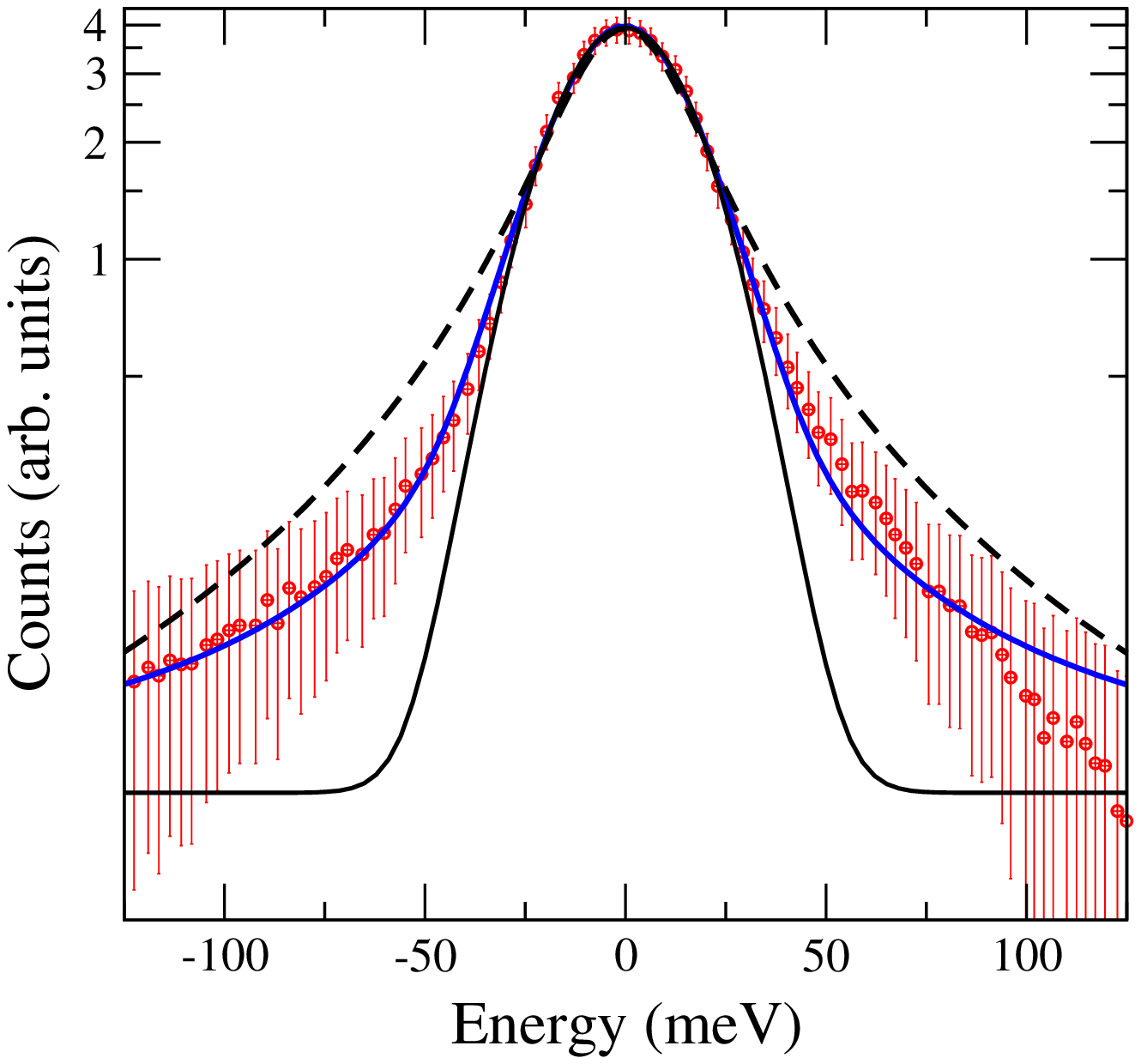}
\caption{
(color online) Resolution curve of the spectrometer. The fit is obtained with a linear combination of a Gaussian (dashed) and a Lorentzian (solid-thin), i.e. pseudo-voigt function (solid-thick). Full width at half maximum (FWHM) is measured to be 38 meV and this number remained unchanged for all incident energies.}
\label{figure1}
\end{figure}

In order to demonstrate the technique, cupric oxide (CuO) was used as a model system. CuO is of particular importance since it is the simplest member of the family that shares the same integral CuO$_4$ plackets with the high-$T_c$ superconducting cuprates \cite{asbrink70}. The CuO single crystal was grown and oriented along the $(100)$ direction \cite{Souptel02} and measured at the 3-ID beamline of the Advanced Photon Source using a RIXS spectrometer based on a sapphire crystal analyzer at a back-scattering geometry \cite{yavas07}. The monochromatic beam of 22 meV resolution was obtained by the four-bounce asymmetrically cut Si (4 4 4) monochromator. The sample was in vacuum to reduce the background due to air scattering, and the measurements were performed at room temperature. The overall energy resolution of the spectrometer was $38$ meV (figure \ref{figure1}) around the Cu $K$-edge as the test measurements of this new spectrometer reported in Ref. \cite{yavas07}. The resolution measurements were repeated for each incident energy before and after the inelastic measurements to assure the observed effect was not due to a possible instrumental glitch. Spectra were collected at four incident photon energies ranging from the $1s\rightarrow 3d$ (quadruple transition) to $1s\rightarrow 4p$ (dipole transition). The incident energy extrema of 8981 eV and 8997 eV were the limits of the instrument as the analyzer was designed to operate at a fixed Bragg angle of around 89.8$^{\circ}$ (very close to back-scattering), and analyzer energy was changed by thermal expansion of the scattering crystal, not the Bragg angle. The temperature of the analyzer crystal, ranging from 100 K to 500 K defines the energy, ranging from 8997 eV to 8981 eV, respectively.

\section{Results and Discussion}

Since there is evidence that electron-lattice coupling is strongest at the zone boundary in cuprates \cite{egami96}, the measurements were taken in the first Brillouin zone such that the momentum transfer vector was along the $(100)$ direction. This geometry corresponds to a scattering angle 2$\theta$ of 16.93$^{\circ}$ with momentum transfer of 13.4 nm$^{-1}$ ($\sim$1.8 nm$^{-1}$ $q$ resolution). Additionally, the phonon branches under investigation have their maximum energies at the zone boundary as reported earlier \cite{Reichardt90}, which makes them relatively easier to detect with the current instrumental resolution.

\begin{figure}
\centering
\includegraphics[scale=0.67]{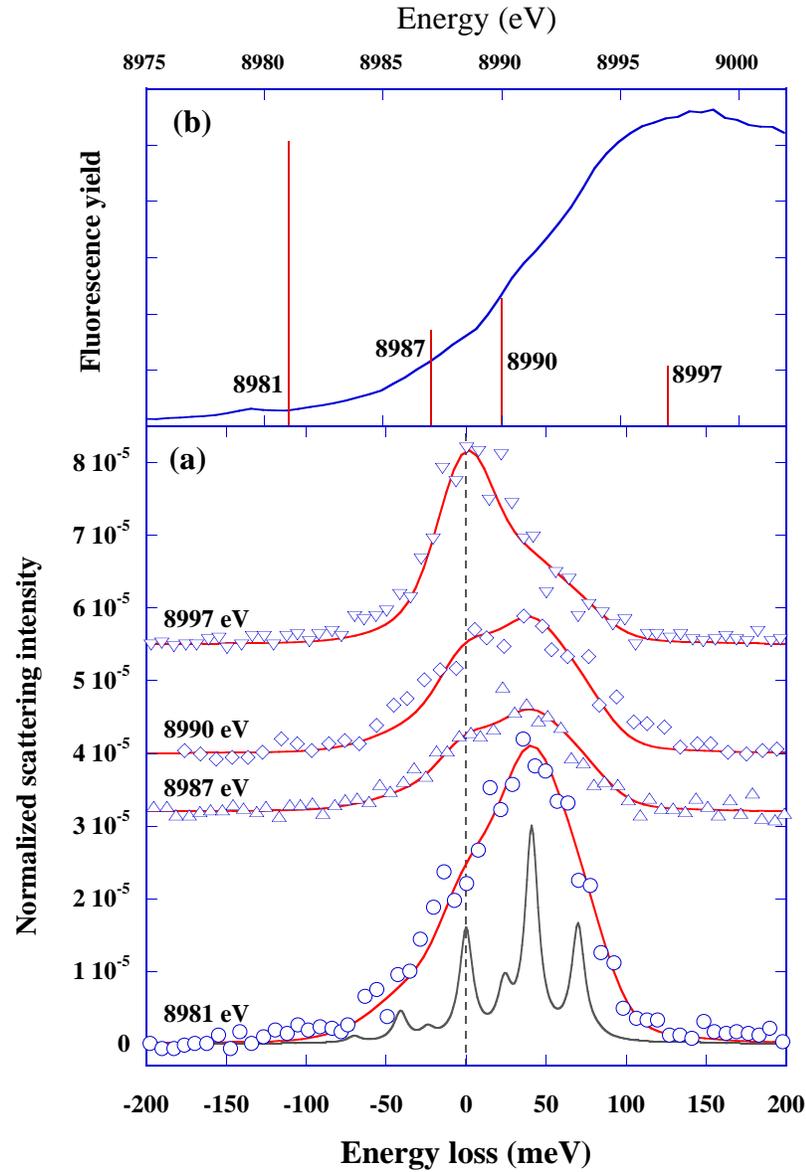}
\caption{
(color online) (a)  Resonant inelastic x-ray scattering spectra at the Cu $K$-edge in CuO for four different incoming energies. The solid lines indicate numerical calculations of the RIXS cross section broadened with a Gaussian with a width of $38$ meV. For $\hbar\omega=8981$ eV, the spectra are shown without additional Gaussian broadening.
 (b) x-ray absorption spectrum in fluorescence yield mode at the Cu $K$-edge in CuO. The lines indicate where the RIXS spectra were taken. The length of the lines indicates the intensity of the inelastic signal.}
\label{figure2}
\end{figure}

The RIXS data are shown in figure \ref{figure2} for several incoming photon energies. In the pre-edge region ($8981$ eV), which is dominated by $1s\rightarrow 3d$ excitations, we observe a clear asymmetry in the spectral features with a maximum intensity around an energy loss of $45$ meV. For larger incoming x-ray energies, the asymmetry remains but the intensity decreases. At an incoming x-ray energy of $8997$ eV, the maximum  coincides with the zero-loss peak with some asymmetry still present. It is known that the elastic signal also enhances as a result of $K$-edge resonance and one can argue that the observed spectral change may be due to this resonant behaviour of the zero-loss peak. However, it was not possible to fit the spectra according to this explanation. The effect of photon absorption coefficients cannot explain the reported anomaly either. The energy scan range is $\mathopen{\pm}$200 meV and the inelastic signal is observed within $\mathopen{\pm}$100 meV. The absorption coefficients are virtually the same for the short energy scanning range compared to the width of the absorption edge of $\sim$20 eV. Therefore, the change in the spectra cannot solely be associated with energy dependence of the absorption coefficients, since photon absorption would only affect the overall intensity, not the line shape. 

It is evident that the inelastic features show a clear resonant behaviour. In RIXS measurements, it is a common practice to take a spectrum away from the transition metal edge to prove the resonant behaviour \cite{kao96,hill98}. However, in this case where the non-resonant measurements away from the Cu $K$-edge are not possible (see above for the discussion about the instrument and ref. \cite{yavas07}), the noticeable change in the spectra as a function of incident photon energy should be accepted as the necessary proof. 

There are several possible explanations for the origin of these inelastic features. We can readily discount crystal-field excitations, since the lowest transition is expected to be on the order of 1-1.5 eV \cite{Eskes90}. Additionally, $K$-edge RIXS is known to be sensitive to magnon excitations \cite{Hill08,Brink07}. Since the RIXS process conserves spin, only two-magnon excitations are allowed with an expected combined energy of 0.3-0.5 eV \cite{Hill08,Brink07}, which is an order of magnitude larger than the features in the present RIXS data. Therefore, we attribute these loss features to the excitations of lattice vibrations. The closeness of the energy loss of the RIXS spectra to the experimentally observed phonon modes supports our assertion \cite{Reichardt90,Guha91}.

\begin{figure}
\centering
\includegraphics[scale=0.80]{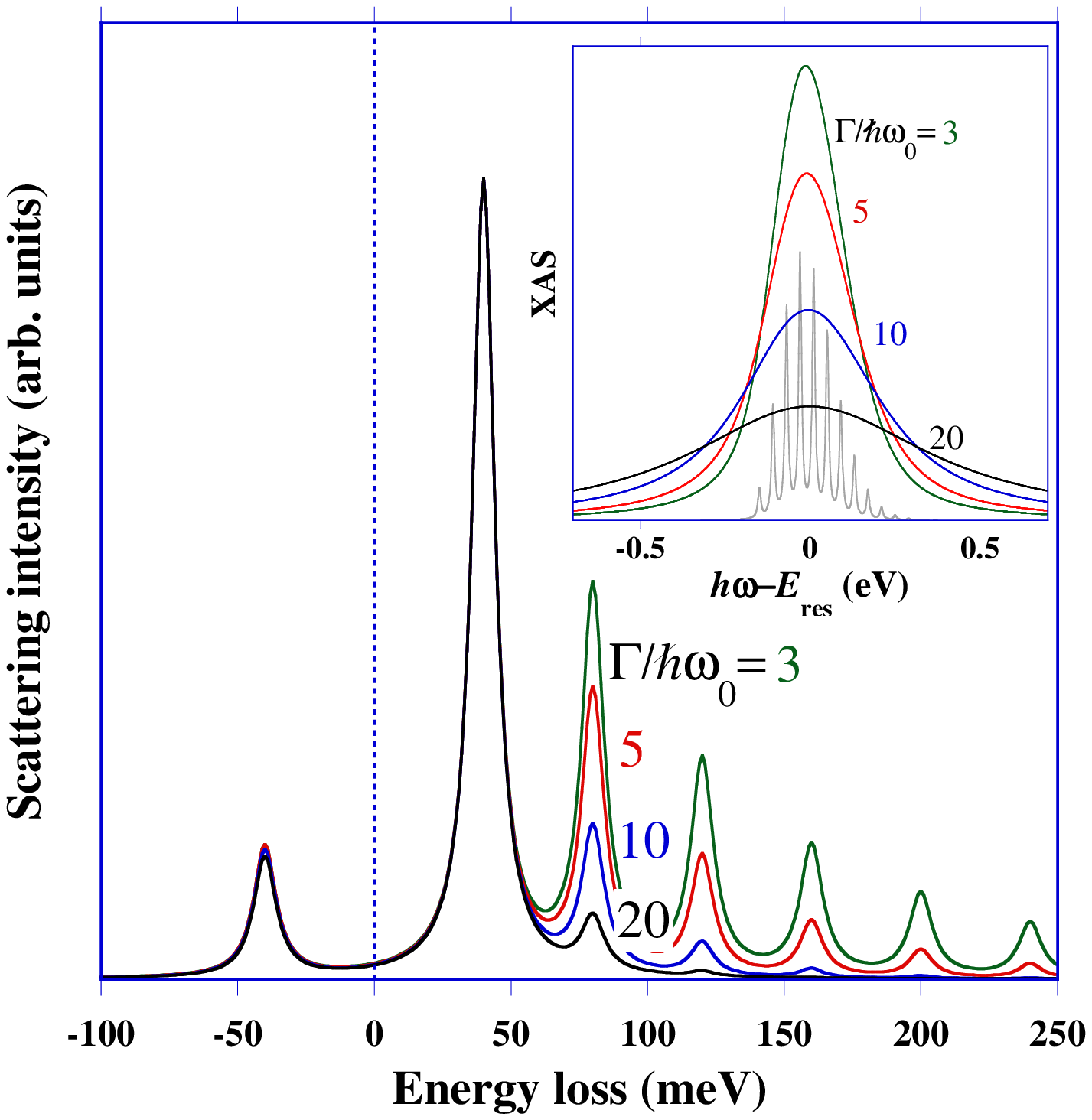}
\caption{
(color online) Exact calculation of the resonant inelastic x-ray scattering for 
a single phonon mode with energy $\hbar\omega_0=40$ meV 
coupled to the transient change in charge distribution on the site
where the absorption takes place (coupling strength $\Delta$ is taken to be $0.15$ eV). 
The calculations clearly show the decrease in intensity of the 
multiphonon excitations with decreasing core-hole lifetime 
(increasing the lifetime broadening $\Gamma$).
The inset
shows the x-ray absorption spectra (XAS). Spectra are given for 
different intermediate-state lifetime broadenings of 
$\Gamma/\hbar\omega_0=$3, 5, 10, and 20. The RIXS spectra are normalized 
to the $40$-meV loss feature.
The grey lines in the inset 
show a XAS spectrum for $\Gamma=\hbar\omega_0/4$, clearly showing the 
phonon excitations.
}
\label{figure3}
\end{figure}

As a result, it can be safely deduced that the observed variation in the spectral shape as a function of incident photon energy can be explained by the change in the relative intensities of the phonon modes against each other. At this geometry, three phonon modes are detectable with energies of 24 meV, 41 meV, and 70 meV \cite{Reichardt90}. We repeated these measurements along the $(100)$ direction to make sure which branches are allowed for this geometry. The rough measurements were done using a 2.2 meV-resolution spectrometer that operates at 21 keV \cite{sinn_nim01}. The experimental data demonstrate the coupling between the lattice and the electronic excitations created by the absorption process and this coupling leads to modulation in the phonon intensities. The electron-phonon interaction can be described by 

\begin{eqnarray}
H=\sum_{{\bf k}m}\hbar\omega_{{\bf k}m}   a^\dagger_{{\bf k}m}a_{{\bf k}m}+
\sum_{{\bf k}m}\sqrt{\Delta_{{\bf k}m} \hbar\omega_{{\bf k}m}} 
 (a^\dagger_{{\bf k}m}+a_{{\bf k}m}),
\end{eqnarray}
where $a^\dagger_{{\bf k}m}$ creates a phonon with mode $m$, wavevector ${\bf k}$, and energy 
$\hbar \omega_{{\bf k}m}$; after the absorption process, the coupling strength to the transient change in charge density with respect to the ground state is determined by  $\Delta_{{\bf k}m}$. In the ground and final states this coupling is zero. This Hamiltonian corresponds to a displaced oscillator, and its solutions are well known. The sudden change in charge density due to the absorption process causes the lattice to respond through the excitation of multiple phonons. To understand the underlying physics, a single dispersionless optical phonon mode with energy 
$\hbar \omega_0$ and coupling $\Delta$ can be taken as an example. In the absorption process, phonons of this mode are excited following a Poisson distribution  with the maximum given by $\Delta/\hbar\omega_0$, see the inset in Fig. \ref{figure3}. For a typical coupling $\Delta$ on the order of $100$ meV, $\Delta/\hbar\omega_0>1$ and it is expected that several phonons are excited in the intermediate state. One therefore expects that several phonons should be visible in the final state. To verify this, the RIXS cross section is calculated using the 
Kramers-Heisenberg equation

\begin{eqnarray}
I(\omega,\omega')=
\sum_{i,f}e^{-\frac{E_i}{k_BT}} \left |\sum_{n}
\frac{ \langle f |{\hat T} |n\rangle \langle n |{\hat T }|i\rangle }{
\hbar \omega+E_i-E_{n} +i\frac{\Gamma}{2}} \right |^2 
\delta(\hbar \omega'+E_f-\hbar \omega -E_i),
\end{eqnarray}
where $|i\rangle$, $|n\rangle$, and $|f\rangle$ are the initial, intermediate, and final states, respectively; 
$\hbar\omega$ and $\hbar\omega'$ are the energies of the incoming and outgoing x-rays; 
${\hat T}\sim {\bf p}\cdot {\bf A}$ is the transition operator; and $\Gamma$ is the full width at half maximum of the lifetime broadening of the core-hole. Figure \ref{figure3} shows that through the Kramers-Heisenberg equation multiple phonon excitations are expected (see the calculations for $\Gamma/\hbar \omega_0=3$ and 5).
Contrary to the calculations, the experimental spectra do not show multiple phonon excitations. This peculiar effect can be explained through numerical calculations by increasing $\Gamma/\hbar \omega_0$. Even though the intermediate states are entirely equivalent in all calculations, the large lifetime broadening $\Gamma$ 
causes a destructive interference between the different intermediate states leading to the same final state. 
To understand this, the RIXS cross section must be more closely examined. The electron-phonon contribution to the intermediate state can be solved by introducing the displaced operators 
${\tilde a}_{{\bf k}m}^\dagger=a_{{\bf k}m}^\dagger+\sqrt{\Delta_{{\bf k}m}/\hbar\omega_{{\bf k}m}}$,
giving a Hamiltonian  
${H}=\sum_{{\bf k}m}{\tilde a}_{{\bf k}m}^\dagger{\tilde a}_{{\bf k}m} \hbar \omega_{{\bf k}m}
-\Delta_{{\bf k}m}$.
Applying an ultra-short core-hole lifetime approximation \cite{Veenendaal96,Brink06,Ament07} to the intermediate-state propagator by expanding the denominator around the resonance energy $E_{\rm res}$ results in
\begin{eqnarray}
\frac{ \langle f |{\hat T} |n\rangle \langle n |{\hat T} |i\rangle }
{z-E_{n}+E_{\rm res}} &=& \sum_{l=0}^\infty
\frac{1}{z^{l+1}}
 \langle f |{\hat T} |n\rangle (E_{n}-E_{\rm res})^l \langle n |
{\hat T} |i\rangle ,
\end{eqnarray}
with $z=\hbar \omega+E_i-E_{\rm res} +i\frac{\Gamma}{2}$. On resonance, where 
$\hbar \omega+E_i\cong E_{\rm res}$ and therefore $|z|\cong \Gamma/2$, the above expression is only valid when
$2|E_n-E_{\rm res}|/\Gamma < 1$. At first, this does not appear to be justified since the eigenstates of ${H}$ go to infinity. However, only the eigenstates with a finite spectral weight, which are spread in an energy region of approximately $2\Delta\ll \Gamma$ around the resonance frequency, are pertinent to the RIXS cross section. Limiting the calculations to the lowest order, an effective scattering amplitude can be obtained as

\begin{eqnarray}
\frac{1}{z^{2}} \langle f |{\hat T}  ({\tilde H}-E_{\rm res}){\hat   T} 
|i\rangle 
=\sum_{{\bf k}m}
\frac{1}{z^2}\sqrt{\Delta_{{\bf k}m} \hbar \omega_{{\bf k}m} }
\langle f |
(a^\dagger_{{\bf k}m} +a_{{\bf k}m} )    |i\rangle ,
\label{amplitude}
\end{eqnarray}
by omitting the terms that do not give rise to inelastic x-ray scattering. The transition operators causing the electronic excitation in the intermediate state effectively cancel due to the unilateral consideration of the phonon excitations in the final state. Therefore in the limit of a very short core-hole lifetime (the large
$\Gamma$ limit), only single phonons will be excited, and the probability will be proportional to the coupling constant. The numerical calculation of the Hamiltonian supports the idea behind the expansion as shown in figure \ref{figure3}. For $\Gamma/\hbar\omega_0=3$ and 5, the lifetime broadening is comparable
to the width of the x-ray absorption spectral features (inset in figure \ref{figure3}); the RIXS spectra clearly show multiple phonon excitations (figure \ref{figure3}) and the restriction to lowest order in the ultrafast core-hole lifetime expansion is invalid. For $\Gamma/\hbar\omega_0=10$ and 20, the multiphonon features are strongly reduced  by the destructive interference, and the spectra are dominated by the single-phonon excitations, described by the amplitude in Equation (\ref{amplitude}). An alternative way to understand the ultra-short core-hole lifetime approximation is by  noting that the decay of the core-hole is so fast that the lattice has insufficient 
time to fully respond to the change in charge distribution caused by the resonant process. At the $K$-edge
in transition metal compounds, where $\Gamma>2$ eV and  $\Delta/\Gamma \ll 1$, the ultra-short core-hole lifetime approximation is well justified. Note that the ratio between the Stokes and anti-Stokes single-phonon 
features remains almost constant for all $\Gamma$. In figure \ref{figure2}, the experimental spectra are compared with a numerical calculation where the intermediate states are exactly diagonalized. The intermediate-state lifetime broadening $\Gamma$ is taken to be $2.0$ eV. Some weight has been added in the zero-loss peak to account for elastic diffuse scattering. The final spectra are further broadened with a Gaussian with a width of $38$ meV to account for experimental uncertainties, including the instrumental resolution. The relative RIXS intensities of the phonon peaks are proportional to $\Delta_{{\bf k}m} \hbar\omega_{{\bf k}m}$ (Equation (1)) and therefore are indicative of the coupling strength between the phonons and the transient change in charge density with respect to the ground state. 

From the intensities of the inelastic phonon peaks, see figure \ref{figure2}(b), we observe a clear resonance behaviour. The largest intensity is observed in the pre-edge region (8981 eV), where direct excitations from 
$1s$ to $3d$ are made. This causes a change in the local $3d$ charge distribution that directly affects the surrounding ligands. The intensity then  drops with the main edge ($>8987$ eV), which is dominated by  excitations into the delocalized $4p$ band that are less effective in coupling to the phonon modes. The multiphonon excitations are strongly suppressed due to the short core-hole lifetime. The best agreement between theory and experiment is obtained when $\Delta_{{\bf k}m}$ for the 24 and 70 meV features are reduced a factor 0.7  compared to the 41 meV phonon excitations. Noting the relative intensities are directly related to the electron-phonon coupling strength, the spectra reveals direct information of the relative corresponding coupling strengths. High-quality data measured on resonance can therefore provide insight into the electron-phonon coupling. The relative ratio between the Stokes and anti-Stokes is given by the Boltzmann factor at room temperature, a further confirmation that the features are phonon related.

In conclusion, we have for the first time demonstrated the resonant enhancement of phonon excitations at the $K$-edge of copper. The experimental features can be explained by the coupling between the phonons and the transient change in the charge distribution on the site where the resonant scattering process occurs. Due to the
fast decay of the core-hole, the system has no time to fully respond to the change in charge density, and 
multiple phonon excitations are strongly suppressed. Using an ultra-fast core-hole expansion, we demonstrate that the resonant inelastic scattering intensity is directly proportional to the electron-phonon coupling strength. Future experiments should include high-$T_c$ superconductors where knowledge of the relative coupling strengths of different phonon modes might shed additional light on their role in superconductivity. We would like to stress that this technique studies the coupling of a specific element (Cu in this case) to the lattice. One can also perform experiments at different edges to obtain element-specific couplings to the phonon modes. Another aspect not explored in this paper is the momentum dependence. The study of the dependence of the intensity of the inelastic phonon peaks as a function of transferred momentum would provide unique insights into the $q$-dependence of the electron-phonon coupling.

\ack
HY acknowledges Thomas Toellner, Ayman Said, Jiyong Zhao, and Daniel Haskel for their help and discussions. MvV was  supported by  the U.S. Department of Energy (DOE), Office of Basic Energy Sciences, Division of Materials Sciences and Engineering under Award DE-FG02-03ER46097. This work was partially supported by COMPRES under NSF Cooperative Agreement EAR 06-49658, and benefited from the RIXS collaboration supported by the Computational Materials Science Network (CMSN) program of the Division of Materials Science and Engineering, Office of Basic Energy Sciences (BES), U.S. DOE under grant number DE-FG02-08ER46540. Use of the Advanced Photon Source was supported by the U.S. DOE, Office of Science, Office of Basic Energy Sciences, under Contract No. DE-AC02-06CH11357.

\pagebreak

\section*{References}
\begin{thebibliography}{30}
\bibitem{lanzara01} Lanzara A {\it et al.} 2001 {\it Nature} {\bf 412} 510 

\bibitem{Cuk04} Cuk T {\it et al.} 2004 {\it Phys. Rev. Lett.} {\bf 93} 117003

\bibitem{Devereaux04} Devereaux T P, Cuk T, Shen Z-X and Nagaosa N 2004 {\it Phys. Rev. Lett.} {\bf 93} 117004
  
\bibitem{reznik06} Reznik D, Pintschovius L, Ito M, Iikubo S, Sato M, Goka H, Fujita M, Yamada K, Gu G D and Tranquada J M 2006 {\it Nature} {\bf 440} 1170

\bibitem{Johnson01} Johnson P D {\it et al.} 2001 {\it Phys. Rev. Lett.} {\bf 87} 177007

\bibitem{giustino08} Giustino F, Cohen M L and Louie S G 2008 {\it Nature} {\bf 452} 975

\bibitem{Dahm09} Dahm T, Hinkov V, Borisenko S V, Kordyuk A A, Zabolotnyy V B, Fink J, Buchner B, Scalapino D J, Hanke W and Keimer B 2009 {\it Nat. Phys.} {\bf 5} 217 

\bibitem{reznik08} Reznik D, Sangiovanni G, Gunnarsson O and Devereaux T P 2008 {\it Nature} {\bf 455} E6

\bibitem{Hancock10} Hancock J N, Chabot-Couture G and Greven M 2010 {\it New J. Phys. } {\bf 12} 033001

\bibitem{Jansen80} Jansen A G M, van Gelder A P and Wyder P 1980 {\it Journal of Physics C: Solid State Physics} {\bf 13} 6073

\bibitem{asbrink70} Asbrink B S and Norrby L J 1970 {\it Acta. Crystallogr.} {\bf B26} 8

\bibitem{Souptel02} Souptel D, Behr G and Balbashov A 2002 {\it Journal of Crystal Growth} {\bf 236} 583

\bibitem{yavas07} Yavas H {\it et al.} 2007 {\it Nucl. Instrum. Methods} {\bf A 582} 149

\bibitem{egami96} Egami T 1996 {\it Journal of Low Temperature Physics} {\bf 105} 791

\bibitem{Reichardt90} Reichardt W, Gompf F,  A\"{i}n M and Wanklyn B M 1990 {\it Zeitschrift f\"{u}r Physik B Condensed Matter} {\bf 81} 19

\bibitem{kao96} Kao C-C, Caliebe W A L, Hastings J B and Gillet J-M 1996 {\it Phys. Rev. B} {\bf 54} 16361

\bibitem{hill98} Hill J P, Kao C-C, Caliebe W A L, Matsubara M, Kotani A, Peng J L and Greene R L 1998 {\it Phys. Rev. Lett.} {\bf 80} 4967

\bibitem{Eskes90} Eskes H, Tjeng L H, and Sawatzky G A 1990 {\it Phys. Rev. B} {\bf 41} 288

\bibitem{Hill08} Hill J P {\it et al.} 2008 {\it Phys. Rev. Lett.} {\bf 100} 097001

\bibitem{Brink07} van~den Brink J 2007 {\it Europhysics Letters} {\bf 80} 47003

\bibitem{Guha91} Guha S, Peebles D and Wieting T J 1991 {\it Phys. Rev. B} {\bf 43} 13092

\bibitem{sinn_nim01} Sinn H {\it et al.} 2001 {\it Nucl. Instrum. Methods} {\bf A 467-468} 1545

\bibitem{Veenendaal96} van Veenendaal M, Carra P and Thole B T 1996 {\it Phys. Rev. B} {\bf 54} 16010

\bibitem{Brink06} van~den Brink J and van Veenendaal M 2006 {\it Europhysics Letters} {\bf 73} 121

\bibitem{Ament07} Ament L J P, Forte F and van~den Brink J 2007 {\it Phys. Rev. B} {\bf 75} 115118

\end{thebibliography}
\end{document}